\documentclass[12pt,journal,epsfig,onecolumn,peerreviewca,draftclsnofoot]{IEEEtran}

\usepackage[dvips]{graphicx}
\usepackage{graphicx}
\usepackage{epstopdf}
\usepackage{amssymb}
\usepackage{cite}
\usepackage{amsmath}

\usepackage{cases}
\usepackage{color}
\usepackage{wrapfig}
\usepackage{url}

\begin{document}

\IEEEoverridecommandlockouts
\title{
An Analytical Range-Angle Dependent Beam Focusing Model for Terahertz Linear Antenna Array
}
\author{ 
Lingxiang Li, Haoran Li, Zhi Chen, ~\IEEEmembership{Senior Member,~IEEE}, \\
 Weixin Chen, Shaoqian Li, ~\IEEEmembership{Fellow,~IEEE}
\thanks{
Lingxiang Li, Haoran Li, Zhi Chen, Weixin Chen and Shaoqian Li are with the National Key Laboratory of Science and Technology on Communications,
UESTC, Chengdu 611731, China (e-mails: \{lingxiang.li, chenzhi\}@uestc.edu.cn).}
\thanks{This work was supported in part by the National Natural Science Foundation of China under Grant U21B2014.}
}

\maketitle

\begin{abstract}
This paper considers a scenario in which the Terahertz (THz) transmitter equipped with a linear antenna array wishes to
focus its beam to a desired spatial region in the array near-field.
The goal is to compute the achievable spatial region and determine how
the system parameters such as the carrier frequency, the array dimension and the user's location affect its beam focusing performance.
First, based on a theorem from analytic geometry, { we show that the achievable focusing spatial region
constitutes a rotated ellipse, with the $x$ and $y$ coordinates denoting the range and angle, respectively.
In this way, the determination of the spatial region is reduced to a problem of deriving the
coverage of an ellipse.}
The achievable coverage is then obtained in
closed form, and the construction of carrier frequency offsets
that can analytically control the beam focusing performance is provided.
Numerical results validate the theoretical findings and demonstrate the performance of the proposed method.
\end{abstract}

\begin{IEEEkeywords}
THz communication, Large-scale antenna arrays, Beam focusing, Near-field, The sixth generation (6G).
\end{IEEEkeywords}


\newtheorem{proposition}{Proposition}
\newtheorem{theorem}{Theorem}
\newtheorem{corollary}{Corollary}
\newtheorem{lemma}{Lemma}

\section{Introduction}\label{SecI}
The sixth generation (6G) wireless systems are expected to support emerging new applications such as augmented reality (AR), visual reality (VR),
and connected autonomous systems \cite{Latif22}. These bandwidth-intensive applications require the delivery hundreds of gigabits per second,
and sensing resolution at the millimeter level. Terahertz (THz) band can provide hundreds of GHz bandwidth and thus is promising in
meeting those requirements \cite{Zhi21,Yinian21}. However, to enjoy the advantage of broad bandwidth of THz in 6G,
some difficulties still need to be resolved, since THz signals suffer from: (i) inherently severe propagation loss; (ii) line-of-sight (LoS) blockage (iii) the effect of molecular absorption noise \cite{Christina22,Hadi21,Zhi19}.

Large-scale antenna arrays and beamforming techniques have received a lot of attention
in compensating for the propagation loss of THz signals and improving the coverage.
One line of research in that area is gearing towards dealing with the beam
squint effect caused by wide bandwidth in THz beamforming systems \cite{Qian21,Jingbo21,Ahmet21},
under the assumption of plane wave propagation model, which applies to far-field scenarios where
the distance between the transmitter and the receiver is greater than or equal to
the Rayleigh distance of the antenna array, i.e., $2D^2/\lambda$ \cite{Frode09}.
Here, $D$ is the maximum dimension of the antenna array, and $\lambda$ denotes the wavelength.
However, this assumption may not hold true for most achievable THz communication or sensing distances. For instance, for an array size of $0.1m$,
the Rayleigh distance is about $4m$ for an operating carrier frequency at 60 GHz. 
Meanwhile, this distance grows to approximately $20m$ at 0.3 THz.
In that case, we refer to the scenario as the \emph{near-field} one, where
spherical wave propagation models should be taken into account.


Although spherical wavefronts have been
extensively studied \cite{Frode09,ZhouZhou15,Emil20}, the literature on \emph{near-field} THz beamforming is relatively sparse \cite{Longfei19,Nitin21}.
The work \cite{Longfei19} follows the line of research in mmWave, and extends the joint two-level
spatial multiplexing and beamforming scheme to THz communications for improving the spectral efficiency in pure LoS conditions.
Different from lower frequencies, because of
the quasi-optical traits of THz wave, the near field beams can focus them at a single focal point by a radially symmetric and linear-ramp field distribution \cite{Daniel18}.
The work \cite{Nitin21} thus considers the beam focus problem for a THz circular planar array, and proposes a kind of
frequency modulated waveform to mitigate the beam misfocus effect caused by wide bandwidth in THz beamforming systems.

To the best of the authors' knowledge, none of existing works discuss how the system parameters affect the beam focusing performance
in the array near-field.
Nevertheless, the spatial focus trait, if properly controlled, is useful in achieving physical-layer (PHY) security for
proximal legitimate user and eavesdropper, and also is useful in reducing electromagnetic interference in networks.
In this work, we consider a scenario in which a THz transmitter wishes to
focus its beam to a desired spatial region.
Our main contributions are summarized as follows.
\begin{enumerate}
\item
We obtain an analytically range-angle dependent beam focusing model for THz linear antenna arrays, uncovering that
{ the achievable focusing spatial region constitutes a rotated ellipse centered at the target, with the $x$ and $y$ coordinates denoting the range and angle, respectively.
\item We determine the achievable coverage of the ellipse in closed form}, as a function of the carrier frequency, the array dimension and the user's location, thus giving insight into how those system parameters affect the array's focusing performance.
\item It shows that due to ultra-short wavelength of THz waves, in the near-field the beampattern transforms
to an ellipse even with a conventional phased array.
Further, we provide two distinct schemes for constructing carrier frequency offsets at antenna elements, through which we can control the beam focusing performance flexibly.
\end{enumerate}

\begin{figure}[t]
\centering
            \includegraphics[scale=0.4]{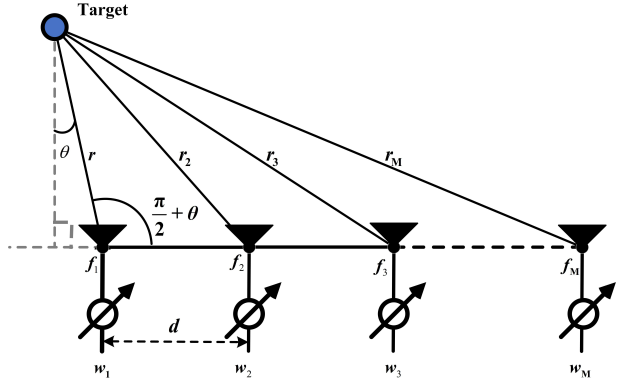}
           \caption{Illustration of the frequency diverse array beamforming approach.}
            \label{system}
        \end{figure}

\section{System Model and Problem Statement}\label{SecII}
As shown in Fig. 1, here we consider a THz transmitter that consists of an { $M$-antenna uniform
array}, with the first antenna element at the origin, and a target user at the location $(r,\theta)$.
By applying cosine rule, the distance from the $m$-th antenna element to the target user is
\begin{align}
r_m&= \sqrt {{r^2} + {{\left( {m - 1} \right)}^2}{d^2} + 2r\left( {m - 1} \right)d\sin \theta }  \nonumber  \\
   &\mathop  \approx \limits^{(a)} r+\frac{(m-1)^2d^2}{2r}+(m-1)d\sin\theta. \label{eq1}
\end{align}
{ Here, the equation (a) holds true under the assumption of ${ (m - 1) d} \ll {r}$,
that is, as compared with the distance $r$ the array size is small enough.}

The transmitter intends to forward information to the target user.
{ As will be shown later in Fig. 2, when the carrier frequency $f_c$ increases to THz, the angle dependent beam transforms to a range-angle dependent ellipse. Besides, noting that the frequency diverse beamforming scheme
has been studied extensively in recent years for far-field radars \cite{WenQin15,Yuan15,Jingran17,Zeeshan21}, wherein results show that the frequency diverse array enables range-dependent radiation patterns.
Therefore, in order to achieve flexible beam focusing in the range dimension,
we { consider} the frequency diverse beamforming approach.
Specifically, a carrier frequency offset $\Delta f_m$ and a weight $w_m$ are employed at the $m$th antenna element, respectively.}

Let $f_c$ be the carrier frequency, the radiation frequency of the $m$th antenna is thus $f_m=f_c+\Delta f_m$.
In addition, the uniform antenna spacing of the transmit array is $d$, which is no less than half of the wavelength to avoid aliasing effects.
{ The signal received at the target user can thus be expressed as
\begin{align}
y\left( {r,\theta ;{\bf f} ,{\bf w}} \right) =   \frac{1}{r}\sum\nolimits_{m = 1}^M {{w _m}{e^{j2\pi f_m {{r _m}/{c}}}}}
\end{align}
where the frequency offset vector ${\bf f}\triangleq [\Delta f_1, \Delta f_2, \cdots, \Delta f_M]$
and ${\bf w}\triangleq [w_1, w_2, \cdots, w_M]$.}

The following assumptions are made in this paper.
\begin{itemize}
\item We assume that the difference in path-loss from antenna elements to the target user can be ignored,
since although the difference in $r_m$s is comparable to the wavelength of THz, it is tiny in absolute value.
\item { We assume that as compared with the distance $r$ the array size is small enough. This assumption
is reasonable for most THz communication scenarios since the wavelength of THz is at the millimeter or even sub-millimeter level.}
\item {  We assume that independent frequency offsets can be applied to a large antenna array.}
\end{itemize}

Same as in phased array, to maximize the antenna gain at the target user at $(R_D,\theta_D)$, at the $m$th antenna we should set the weight
${ {w _m^\star}}={\text {exp}}(-j2\pi {f_m}r_m^D/c)$, where $r_m^D$ is obtained by setting $(r, \theta)$ in (\ref{eq1}) as $(R_D,\theta_D)$.
The resulting beampattern can thus be computed as,
\begin{align}
S\left( {r,\theta ;{\bf f}} \right) =  &  {\left|\sum\nolimits_{m = 1}^M {{e^{j2\pi f_m {(r_m-r_m^D)}/{c}}}} \right|^2}.
 \end{align}


Generally, the determination of the beampattern for near-field is a non-convex problem.
For the purpose of exhibiting the results in a provable way, we
introduce the half-power points, i.e. the 3dB main lobe, as an alternative performance metric.
That is, $(R_B, \theta_B)$ are those points { where the received power reduces to half of that
received by the target user}, i.e.,
\begin{align}
\mathcal{B}  \buildrel \Delta \over = \{ (R_B, \theta_B)| S\left( {R_B,\theta_B ;{\bf f}} \right) = M^2/2\}. \label{eq4}
\end{align}

From (\ref{eq4}), one can see that the achievable boundary of $\mathcal{B}$
not only depends on the position of the target user, but also
depends on the frequency offsets at the transmit antenna elements.
In order to focus the transmitting energy on any desired spatial region,
we need to find out how those factors affect the boundary of $\mathcal{B}$.
In the following, we will first determine $\mathcal{B}$ in closed form,
and then giving two strategies to control the beampattern
by adjusting the frequency offsets at transmit antenna elements.

\section{The Range-Angle Dependent Beam Focusing Model for THz Linear Antenna Array}\label{SecIII}
In this section, we will derive the beampattern in closed-form, thus giving a beam focusing model and a way to analytically evaluate the beam focusing performance.
The key idea is to reformulate the beampattern expression in (\ref{eq4}) by employing theorem of analytic geometry (see \emph{Lemma \ref{lem1}}),
with which we show that the achievable spatial region
constitutes { a rotated ellipse centered at $(R_D, \theta_D)$, with the $x$ and $y$ coordinates denoting $R_B$ and $\theta_B$, respectively.}
As such, the determination of the spatial region is reduced to a problem of deriving
{  the coverage of a rotated ellipse. Then by \emph{Theorem \ref{theom1}} we determine the achievable coverage of the rotated ellipse along the range dimension and angular dimension in
closed form.} Details are shown in the following text.

\begin{lemma}\label{lem1}
\textit{For a THz linear array, the boundary of $\mathcal{B}$ given
in (\ref{eq4}) satisfies the following equation:}
\begin{align}
&X{\left( {{R_B} - {R_D}} \right)^2} + 2Y\left( {{R_B} - {R_D}} \right)\left( {{\theta _B} - {\theta _D}} \right) +  \nonumber \\
&Z{\left( {{\theta _B} - {\theta _D}} \right)^2} - {M^2} = 0, \label{eqLemma1}
\end{align}
where $X$, $Y$ and $Z$ are given as follows.
\begin{align}
X = &\frac{2{{\pi ^2}}}{{{c^2}}}\sum\nolimits_{m = 1}^M {\sum\nolimits_{n = 1}^M {{{\left[ {2\left( {{\xi _m} - {\xi _n}} \right)} \right]}^2}} } \nonumber \\
Y = &\frac{{4{\pi ^2}d f_c\cos {\theta _D}}}{{{c^2}}}\sum\nolimits_{m = 1}^M {\sum\nolimits_{n = 1}^M {\left[ {2\left( {{\xi _m} - {\xi _n}} \right)\left( {m - n} \right)} \right]} }\nonumber \\
Z = &\frac{{8{\pi ^2}{f_c^2}{d^2}{{\cos }^2}{\theta _D}}}{{{c^2}}}\sum\nolimits_{m = 1}^M {\sum\nolimits_{n = 1}^M {{{\left( {m - n} \right)}^2}} }, \label{eq6}
\end{align}
with ${\xi _m} =  {\Delta f_m} - \frac{{{d^2}f_c}}{{2R^2_D}}{\left( {m - 1} \right)^2}, \forall m$.
\end{lemma}
\begin{IEEEproof}
See Appendix \ref{appA}.
\end{IEEEproof}

Based on \emph{Lemma \ref{lem1}}, one can see that the type of the half-power boundary is depended on the value of
$X$, $Y$, and $Z$. From the geometry knowledge, it holds true that, if $XZ > {Y^2}$ the half-power boundary is an ellipse. Otherwise,
the half-power boundary is a hyperbola or a pair of parallels for the case of $XZ < {Y^2}$ and $XZ = {Y^2}$, respectively. For the last two cases,
it is impossible to achieve beam alignment in reality, which is of no interest to the communication community. Thus, in this paper we only
focus on the first case where the half-power boundary is an ellipse.

\begin{theorem}\label{theom1}
\textit{For a THz beam focusing system with linear antenna array,
the achievable ellipse area, the main lobe beamwidth in the range and
angle dimension are respectively, }
\begin{align}
S &=  \frac{{\pi {M^2}}}{{\sqrt {XZ - {Y^2}} }}, \nonumber \\
\Delta_R & \triangleq \max(R_B)-\min(R_B) = 2\sqrt {\frac{{{M^2}Z}}{XZ - Y^2}},  \nonumber \\
\Delta_\theta & \triangleq \max(\theta_B)-\min(\theta_B) = 2\sqrt {\frac{{{M^2}X}}{XZ - Y^2}}. \nonumber
\end{align}
\end{theorem}
\begin{IEEEproof}
See Appendix \ref{appB}.
\end{IEEEproof}

Specially, for the case without frequency offsets at antenna elements, i.e., $\Delta{f_m} = 0$, $\forall m$,
we get a conventional phased array. Substituting $\Delta{f_m} = 0$, $\forall m$
into \emph{Theorem 1}, we arrive at \emph{Corollary \ref{Coro1}} as follows.

\begin{corollary}\label{Coro1}
\textit{Provided that $\Delta f_m=0, \forall m$, the frequency diverse array degenerates to a
phased array, and the ellipse area, the main lobe beamwidth in the range and
angle dimension are respectively,}
\begin{align}
S =  &   \frac{{3\sqrt {15} {c^2}R_D^2}}{{\pi f_c^2{d^3}\cos {\theta _D}({M^2} - 1)\sqrt {{M^2} - 4} }} \nonumber \\
{{\Delta }_R} =  &    \frac{{6\sqrt {10} cR_D^2}}{{\pi f_c^{}{d^2}\sqrt {({M^2} - 1)\left( {{M^2} - 4} \right)} }} \nonumber \\
{{\Delta}_\theta } =  &  \frac{{c\sqrt {6\left( {16{M^2} - 30M + 11} \right)} }}{{\pi f_c^{}d\cos {\theta _D}\sqrt {({M^2} - 1)\left( {{M^2} - 4} \right)} }} \nonumber
\end{align}
\end{corollary}

\emph{Remark 1:} According to \emph{Corollary 1}, one can see that the ellipse area of a phased array decreases as the number of antennas $M$ increases,
which indicates that as the carrier frequency $f_c$ increases to THz, the angular dependent beam transforms to a range-angle dependent ellipse.
This coincides with the numerical results as shown in Fig. 2, where the array antenna size $D=0.3m$,
and a target user is at the location of $(R_D,\theta_D)=\left( {10m,{{20}^ \circ }} \right)$.

\emph{Remark 2:} The main lobe beamwidth in the range dimension increases as the distance $R_D$ from the transmitter increases,
which indicates that as the target user moves from the near-field to the far-field range, the beam focusing ability declines.
Furthermore, the main lobe beamwidth in the angular dimension is independent of the distance $R_D$.

\emph{Remark 3:} The property of ultra-short wavelength of THz wave can be a double-edged sword. On the one hand, as compared to lower frequencies,
THz phased array gets higher resolution and can focus RF signals to a two-dimensional spatial region instead of only along the angular direction. One the other hand, THz signals suffer from
limited range coverage, which challenges the conventional beam scanning and tracking methods that only differentiate beams with respect to angular dimension.
\begin{figure}[t]
\centering
  \includegraphics[width=\linewidth]{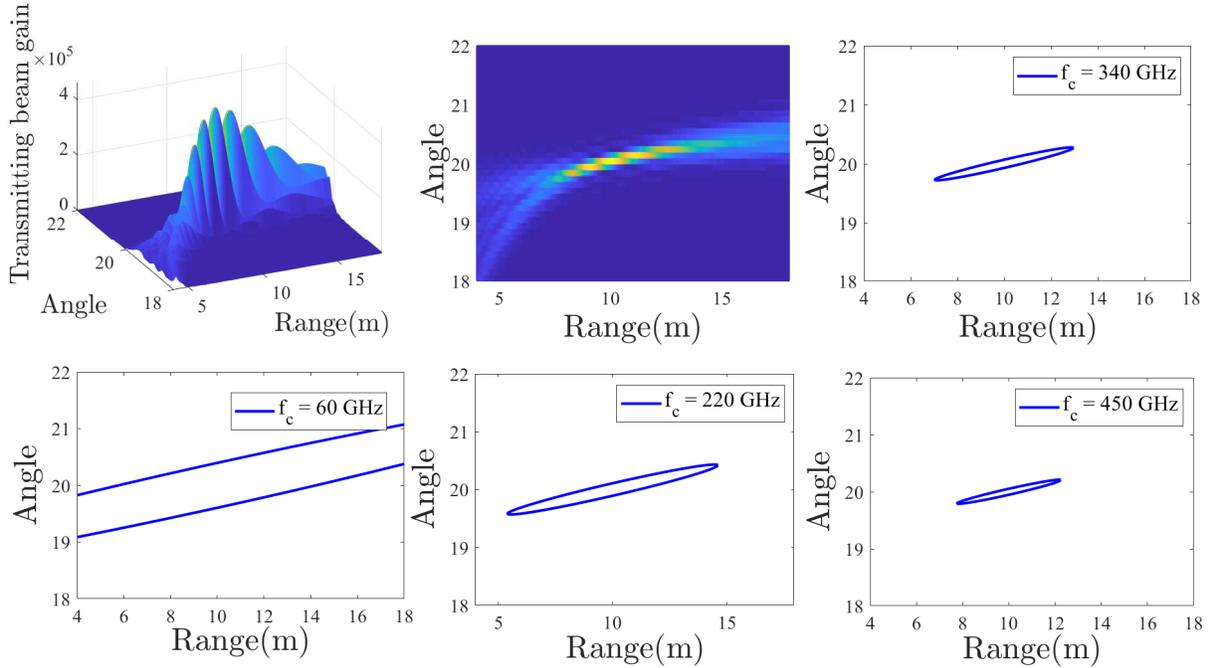}
  \caption{Beampattern of the phased array: transformation from beam to ellipse.}
  \label{system}
\end{figure}

\section{Proposed Transmission Schemes that can Focus Beam Flexibly}\label{SecIII-A}
{ By {\textit {Theorem}} 1, we show that the achievable spatial region
constitutes an ellipse, with the $x$ and $y$ coordinates denoting $R_B$ and $\theta_B$, respectively.
Furthermore, the achievable coverage of the ellipse depends on system parameters such
as the carrier frequency, the array dimension and the user's location.}
In this section, we give two distinct schemes for constructing carrier frequency offsets at antenna elements, through which the beam focusing performance is flexibly controlled.

As aforementioned, we only need to focus on the case where the half-power boundary is an ellipse and
$XZ > {Y^2}$, which holds true if and only if
\begin{align}
\Delta {f_m} \ne \frac{{\beta m}}{2} + \frac{{f_c{d^2}}}{{2R_D^2}}{\left( {m - 1} \right)^2} + C,  \label{eq7}
\end{align}
where $\beta$ and $C$ are constants.
In other words, to make sure the energy is delivered to an intended two-dimensional spatial region, the only
prerequisite is that the equation (\ref {eq7}) holds true. Meanwhile, the specific value of
$\Delta {f_m}$s decides the values of $\Delta_R$ and $\Delta_\theta$.

\begin{proposition}\label{prop1}
\textit{Let $\Delta f_m=\dfrac{{f_c{d^2}}}{{2R^2_D}}\alpha {\left( {m - 1} \right)^2}, \forall m $, then the main lobe beamwidth in the range and angle dimension are respectively}
\begin{align}
{\Delta _R} =  & \dfrac{2}{{ { |1-\alpha|} }}\sqrt {\frac{{{M^2}{Z_{{\rm{pa}}}}}}{{{X_{{\rm{pa}}}}{Z_{{\rm{pa}}}} - Y_{{\rm{pa}}}^2}}} \nonumber \\
{\Delta _\theta } =  & 2\sqrt {\frac{{{M^2}{X_{{\rm{pa}}}}}}{{{X_{{\rm{pa}}}}{Z_{{\rm{pa}}}} - Y_{{\rm{pa}}}^2}}} \nonumber
\end{align}
\textit{where $\alpha$ is an undermined parameter. $X_{{\rm{pa}}}$, $Y_{{\rm{pa}}}$ and $Z_{{\rm{pa}}}$ equal $X$, $Y$ and $Z$, respectively, by setting the frequency offsets in (\ref {eq6}) as zero.}
\end{proposition}

\begin{proposition}\label{prop2}
\textit{Let $\Delta f_m= \frac{\delta }{2}\left|{\frac{{\sin (m - 1)}}{\pi }} \right| + \frac{{f_c{d^2}}}{{2R_D^2}}{(m - 1)^2}$, $\forall m$, then the main lobe beamwidth in the range and angle dimension are respectively}
\begin{align}
{\Delta _R} =  & \frac{2}{\delta }\sqrt {\frac{{{M^2}{Z_{\rm fa}}}}{{{X_{\rm fa}}{Z_{\rm fa}} - Y_{\rm fa}^2}}}\nonumber \\
{\Delta _\theta } =  & 2\sqrt {\frac{{{M^2}{X_{\rm fa}}}}{{{X_{\rm fa}}{Z_{\rm fa}} - Y_{\rm fa}^2}}} \nonumber
\end{align}
\textit{where $\delta$ is an undermined parameter. $X_{\rm fa}$, $Y_{\rm fa}$ and $Z_{\rm fa}$ equal $X$, $Y$ and $Z$, respectively, by setting $\xi_m=\left|{\frac{{\sin (m - 1)}}{\pi }} \right|$ in (\ref {eq6}).}
\end{proposition}

Substituting $\Delta f_m$ into \emph{Theorem 1} and by some mathematical reformulations, we get the
conclusions as in \textit{Proposition} 1 and \textit{Proposition} 2. This completes the proof.

It is obvious that for any intended range of ${\Delta _R}= \rho$, we
can derive the specific value of $\alpha$, $\delta$, and frequency offsets by substituting ${\Delta _R}= \rho$
into \textit{Proposition} 1 and \textit{Proposition} 2, respectively.
The main lobe beamwidth in range dimension $\Delta _R$ decreases monotonically with $|1-\alpha|$ and $\delta$, respectively. Meanwhile, the main lobe beamwidth in angular dimension $\Delta _\theta$ remains a constant.



For example, let the carrier frequency be 340GHz, the array
antenna size $D=0.3m$, and a target user is at the location of $(R_D,\theta_D)=\left( {15m,{{20}^ \circ }} \right)$.
By adjusting the parameters $\alpha$ and $\delta$, we get different frequency offsets which are shown in Table 1 and Table 2, respectively.
As a result, as shown in Fig. 3, the beam focusing performance is flexibly controlled.

\medskip
\begin{table} [h]
\centering
\vspace{-0.4cm}   
\caption{Proposed scheme 1 with different $\alpha$}
\label{tab1}
\vspace{-0.2cm}
\resizebox{\columnwidth}{!}{
\begin{tabular}{|l|c|c|c|c|c|c|c|}
  \hline
   $\alpha$& $m=1$ & $m=2$ & $m=3$ & $\cdots$ &  $m=679$ & $m=680$ \\
  \hline
  $0.2$ & 0 & 29.4Hz  & 117.6Hz & $\cdots$ &  13.5MHz& 13.6MHz \\
  \hline
  $0.4$ & 0 & 58.8Hz  & 235.3Hz & $\cdots$ & 30.0MHz & 27.1MHz \\
  \hline
  $0.6$ & 0 & 88.3Hz  & 352.9Hz & $\cdots$ &  40.5MHz & 40.7MHz \\
  \hline
  $0.8$ & 0 & 117.6Hz & 470.6Hz & $\cdots$ & 54.1MHz & 54.2MHz \\
  \hline
\end{tabular}
}

\end{table}

\begin{table} [h]
\centering
\vspace{-0.6cm}   
\caption{Proposed scheme 2 with different $\delta$}
\label{tab1}
\vspace{-0.2cm}   
\resizebox{\columnwidth}{!}{
\begin{tabular}{|c|c|c|c|c|c|c|c|}
  \hline
  $\delta $ & $m=1$ & $m=2$ & $m=3$ & $\cdots$ &  $m=679$ & $m=680$ \\
  \hline
  $6 \times {10^7}$  & 0 & 8.0MHz   & 8.7MHz  & $\cdots$ &  72.9MHz& 71.2MHz \\
  \hline
  $12 \times {10^7}$ & 0 & 16.1MHz  & 17.4MHz & $\cdots$ & 78.1MHz & 75.5MHz \\
  \hline
  $18 \times {10^7}$ & 0 & 24.1MHz  & 26.1MHz & $\cdots$ &  83.4MHz & 79.3MHz \\
  \hline
  $24 \times {10^7}$ & 0 & 32.1MHz  & 34.7MHz & $\cdots$ & 88.7MHz & 83.2MHz \\
  \hline
  $30 \times {10^7}$ & 0 & 40.1MHz  & 43.4MHz & $\cdots$ & 93.9MHz & 87.1MHz \\
  \hline

\end{tabular}
 }
\end{table}

{ \textit{Remark 4:}
It is worthwhile to note that there are many other feasible solutions to
control the beam focusing performance. Due to the limited space,
we only enumerate two of them and show that it is possible
to flexibly focus the beam to a two-dimensional spatial region
with a THz linear antenna array.
In {\textit{Proposition 1}} the frequency offsets over different antennas are integer times of a fundamental value, i.e.,
${{f_c{d^2}}\alpha}/{{2R^2_D}}$. In comparison, {\textit{Proposition 2}} needs more fine-grained frequency offsets over different antennas, which varies with the term ${{\sin (m - 1)}}/{\pi }$, with $m$ denoting the antenna index.
On the other hand, current THz sources are produced with frequency multipliers,
which only support coarse-grained frequency offsets.
Therefore, from the perspective of implementation, the realization of {\textit{Proposition 2}} needs digital beamforming,
which requires higher power consumption and higher system cost. }

\begin{figure}[t]
\centering
  \includegraphics[width=\linewidth]{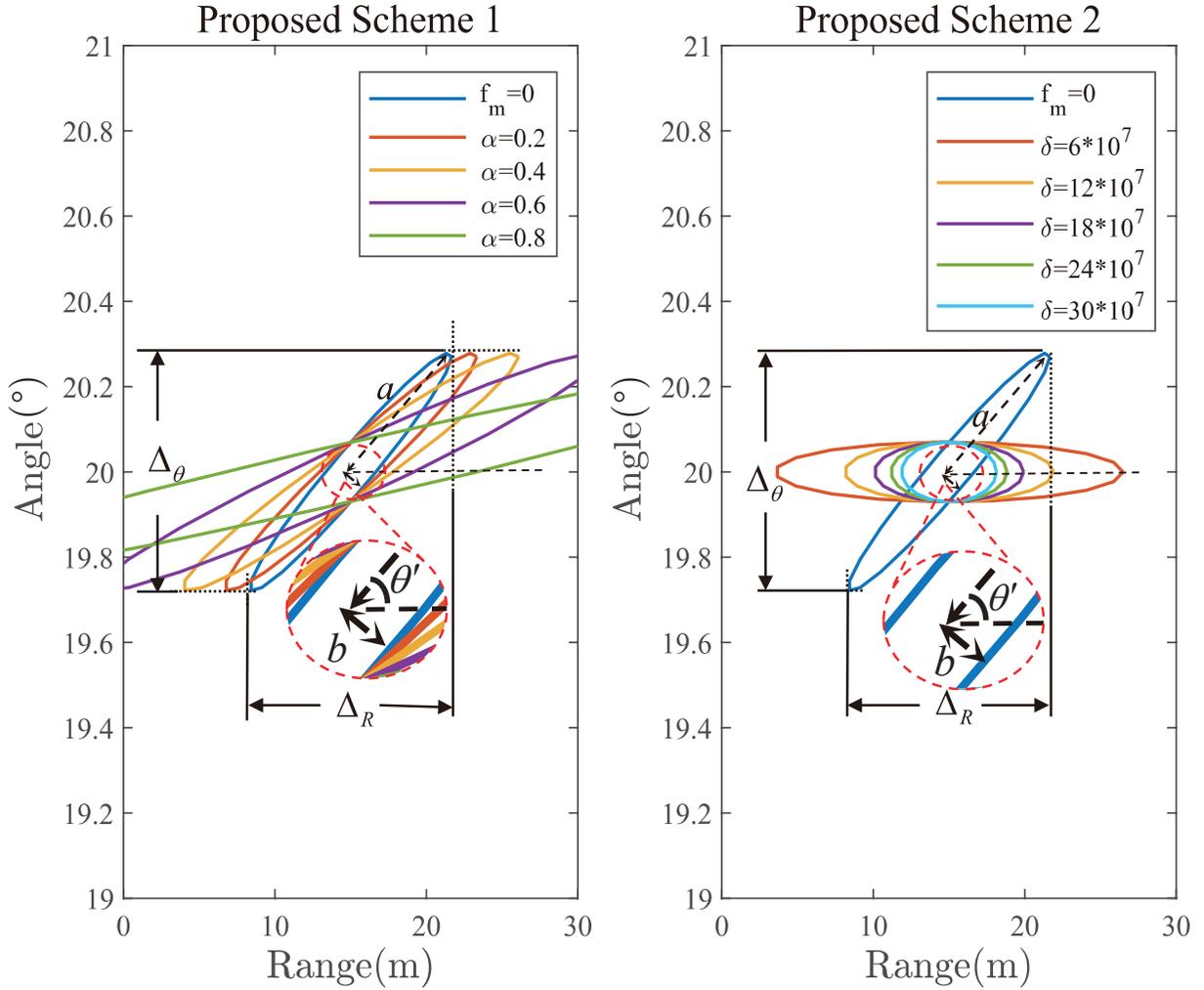}
  \caption{Beampattern obtained with different parameters $\alpha$ or $\delta$.}
  \label{system}
\end{figure}

\section{Conclusion}
We have analytically addressed the beam focusing problem
of a THz transmitter which wishes to
focus its beam to a desired spatial region in the array near-field.
Specifically, based on a theorem from analytic geometry, we have shown that the achievable spatial region
constitutes an ellipse. 
We have further determined in closed form the achievable coverage of the ellipse, revealing
how the system parameters such as the carrier frequency, the array dimension and the user's location affect its focusing performance.
We have also given two schemes to analytically control the coverage of the ellipse by adjusting carrier frequency offsets.
Numerical results have validated the theoretical findings and demonstrate the performance of the proposed method.

\appendices

\section{Proof of \emph{Lemma 1}} \label{appA}
According to (\ref{eq4}), the half-power boundary points of $\left( {{R_B},{\theta _B}} \right)$ satisfies
the following equation:
\begin{align}
\frac{{{M^2}}}{2} =& \left|{\sum\limits_{m = 1}^M {\exp \left( {j\frac{{2\pi f_m\left( {r_m^B - r_m^D} \right)}}{c}  } \right)} }\right|^2, \nonumber \\
\mathop  = \limits^{(a)} & \left( {\sum\nolimits_{m = 1}^M {{e^{j{x_m}}}} } \right){\left( {\sum\nolimits_{m = 1}^M {{e^{j{x_m}}}} } \right)^ * } \nonumber \\
\mathop  = \limits^{(b)} &\sum\nolimits_{n = 1}^M {\sum\nolimits_{m = 1}^M {\cos \left( {{x_m} - {x_n}} \right)} }, \nonumber \\
\mathop  \approx \limits^{(c)} &\sum\nolimits_{n = 1}^M {\sum\nolimits_{m = 1}^M {(1 - \frac{1}{2}{\left( {{x_m} - {x_n}} \right)^2} )} }, \label{eqA1}
\end{align}
where (a) is obtained by letting ${x_m}={{2\pi f_m\left( {r_m^B - r_m^D} \right)}}/{c}$,
with $r_m^B$ obtained by setting $(r, \theta)$ in (\ref{eq1}) as $(R_B,\theta_B)$;
(b) holds true due to using Euler's Theorem;
{  (c) is obtained by using the approximation 
${x_m} \approx {x_n}$. This approximation is usually valid in
THz near-field scenarios, where due to the quasi-optical traits of THz waves the half-power boundary points are assumed to be near the maximum power point.}
Therefore, the equation (\ref{eqA1}) can be re-expressed as
\begin{align}
\sum\nolimits_{m = 1}^M {\sum\nolimits_{m = 1}^M {{{\left( {{x_m} - {x_n}} \right)}^2}} }  - {M^2} = 0. \label{eqb3}
\end{align}

On the other hand, looking into the formula of ${\left( {{x_m} - {x_n}} \right)}^2$, we have the following derivations,
\begin{align}
&{\left( {{x_m} - {x_n}} \right)^2}
 =   \frac{{4{\pi ^2}}}{{{c^2}}}\left\{ { {f_m}} \right.\left( {r_m^B - {r_m^D}} \right) - {\left. { {{f_n}} \left( {r_n^B - {r_n^D}} \right)} \right\}^2}\nonumber\\
\mathop  = \limits^{(a)}  & \frac{{4{\pi ^2}}}{{{c^2}}}\left\{ {\left[ {{f_c}\left( {r_m^B - {r_m^D}} \right) + \Delta {f_m}\left( {{R_B} - {R_D}} \right)} \right]} \right.\nonumber\\
 &  - {\left. {\left[ {{f_c}\left( {r_n^B - {r_n^D}} \right) + \Delta {f_n}\left( {{R_B} - {R_D}} \right)} \right]} \right\}^2}\nonumber\\
\mathop  = \limits^{(b)}  & \frac{{2{\pi ^2}}}{{{c^2}}}{\left[ {2\left( {{\xi _m} - {\xi _n}} \right)} \right]^2}{\left( {{R_B} - {R_D}} \right)^2}\nonumber\\
 &  + \frac{{2 \times 4{\pi ^2}d{f_c}\cos {\theta _D}}}{{{c^2}}}\left[ {2\left( {{\xi _m} - {\xi _n}} \right)\left( {m - n} \right)} \right]\nonumber\\
 &\left( {{R_B} - {R_D}} \right)\left( {{\theta _B} - {\theta _D}} \right) & \nonumber\\
 &  + \frac{{8{\pi ^2}{f_c^2}{d^2}{{\cos }^2}{\theta _D}}}{{{c^2}}}\left( {m - n} \right)\left( {{\theta _B} - {\theta _D}} \right)^2, \label{eqb5}
\end{align}
where ${\xi _m} \triangleq  {\Delta f_m} - \frac{{{d^2}f_c}}{{2R^2_D}}{\left( {m - 1} \right)^2}, \forall m $.
{ The equation (a) is obtained by substituting (\ref{eq1}) into $( {{x_m} - {x_n}} )^2$, and applying
the facts that ${{2\pi \Delta {f_m}\left( {m - 1} \right){d^2}} \mathord{\left/
 {\vphantom {{2\pi \Delta {f_m}\left( {m - 1} \right){d^2}} {2r}}} \right.
 \kern-\nulldelimiterspace} {2r}} $ and $2\pi \Delta {f_m}\left( {m - 1} \right)d\sin \theta  $
 are small enough and can be omitted.}
The equation (b) is obtained by letting $\left( {\sin {\theta _B} - \sin {\theta _D}} \right) \approx  {\cos {\theta _D}} \left( {{\theta _B} - {\theta _D}} \right)$,
and executing the second-order Taylor expansion around $\left( {{R_D},{\theta _D}} \right)$, that is, $\left( {{R_B},{\theta _B}} \right)$.
This can be done since the half-power boundary points are near the target user.

Substituting (\ref{eqb5}) into (\ref{eqb3}) and doing some reformulations, we arrive at conclusions in
\emph{Lemma 1}. This completes the proof.

\section{Proof of \emph{Theorem 1}} \label{appB}
{ According to the theorem of analytic geometry{\cite{Leon2015}}, if an equation has the form given in (\ref{eqLemma1}),
it must correspond to a rotated ellipse centered at $(R_D, \theta_D)$, with the $x$ and $y$ coordinates denoting $R_B$ and $\theta_B$, respectively.
Its area, semi-major and semi-minor axes are, respectively,
\begin{subequations}
\begin{align}
S &= \pi ab = {{\pi {M^2}}}/{{\sqrt {XZ - {Y^2}} }},  \\
a &=\sqrt {{{2{M^2}}}/{{\left( {X + Z}  - \sqrt {{{\left( {X - Z} \right)}^2} + 4{Y^2}}\right) }}},  \\
b &=\sqrt {{{2{M^2}}}/{{\left( {X + Z}  + \sqrt {{{\left( {X - Z} \right)}^2} + 4{Y^2}}\right) }}}.
\end{align}
\end{subequations}
Moreover, the rotation angle of the ellipse from
the $x$-axis, denoted by ${\theta '}$, satisfies
\begin{align}
{\cos ^2}{{\theta '}} = {1}/{2} + {(Z - X)}/{{2 \sqrt {{{\left( {X - Z} \right)}^2} + 4{Y^2}} }}. \nonumber
\end{align}
The coordinates of the boundary point on the ellipse can thus be expressed as \cite{Jong2013},
\begin{subequations}
\begin{align}
{R_B} &=   a\cos \theta '\cos t - b\sin \theta '\sin t + {R_D},   \\
{\theta _B} &= a\sin \theta '\cos t + b\cos \theta '\sin t + {\theta _D},
\end{align}
\end{subequations}
where $t \in \left[ {0,2\pi } \right]$. }

Therefore, according to the auxiliary angle formula \cite{Barry2001},
the coverage of the ellipse in the range dimension and angle dimension are, respectively,
\begin{align}
{\Delta _R} &= \max \left( {{R_B}} \right) - \min \left( {{R_B}} \right) = 2\sqrt {{a^2}{{\cos }^2}\theta ' + {b^2}{{\sin }^2}\theta '}   \nonumber \\
&=  2\sqrt {{{{M^2}Z}}/(XZ - {Y^2})}, \label{eqb6}
\end{align}
\begin{align}
{\Delta _\theta } &=  \max \left( {{\theta _B}} \right) - \min \left( {{\theta _B}} \right) = 2\sqrt {{a^2}{{\sin }^2}\theta ' + {b^2}{{\cos }^2}\theta '} \nonumber \\
&=  2\sqrt {{{{M^2}X}}/{(XZ - {Y^2})}}.\label{eqb7}
\end{align}

Note that $X$, $Y$ and $Z$ are given by (\ref{eq6}).
This completes the proof.

\bibliography{mybib}
\bibliographystyle{IEEEtran}

%

\end{document}